\documentclass[12pt, amsmath]{iopart}

 \pdfoutput=1

\newcommand{\defeq}{:=}

\usepackage{hyperref}
\usepackage{graphicx}

\graphicspath{{figs/}}

\date{\today}

\newcommand{\eqref}[1]{(\ref{#1})}

\newcommand{\cc}[3]{c_{#1}(#2, #3)}

\newcommand{\tp}{{p}}
\newcommand{\tq}{{q}}

\date{\today}

\newcommand{\Jhat}{\hat{J}}
\newcommand{\Chat}{\hat{C}}
\newcommand{\What}{\hat{W}}

\usepackage[utf8]{inputenc}

\begin{document}

\title[Emergence of a current by coupling equilibrating systems]{Emergence of a sustained current by coupling equilibrating systems: Making a NESS out of equilibrium} 

%\author{Hern\'an \surname{Larralde}}
%\author{Fran\c{c}ois \surname{Leyvraz}}
%\author{David P.\ \surname{Sanders}}

\author{H Larralde$^1$ and DP Sanders$^2$} 
\address{$^1$ Instituto de Ciencias
F\'{\i}sicas, Universidad Nacional Autónoma de México, Apartado postal 48-3, C.P.\ 62551, Cuernavaca,
Morelos, Mexico\\
$^2$ Departamento de F\'{\i}sica, Facultad de Ciencias, Universidad Nacional Autónoma de México, Ciudad Universitaria,
Del.~Coyoacán, México D.F.~04510, Mexico}
\eads{\mailto{hernan@fis.unam.mx} and \mailto{dpsanders@ciencias.unam.mx}}
\date{\today}

\begin{abstract}
We show that coupling together two closed thermodynamic systems that independently attain equilibrium
may give rise to a nonequilibrium stationary state (NESS) 
with a persistent, non-vanishing current. We study a simple example that is exactly soluble, consisting of
random walkers with different biases towards a reflecting boundary,
modelling, for example, Brownian particles with different charge states in an electric field.
% modeling particles with different charge states in an electric field.
We obtain exact analytical expressions for the
(generating functions of)
concentrations and currents in the NESS for this model, and exhibit the main features by
numerical simulation.
\end{abstract}
% 
% 
% We show that it is possible to construct a nonequilibrium stationary state (NESS) in a closed
% system with a persistent, non-vanishing current, by
% coupling two simple systems, each of which  independently attains
% equilibrium.  These systems are composed of random walkers on
% one-dimensional semi-infinite lattices with an overall bias towards a
% reflecting boundary. We obtain exact analytical expressions for the
% generating functions of various relevant quantities, including the
% concentrations and currents in the NESS
%}
%\pacs{05.70.Ln}{Nonequilibrium and irreversible thermodynamics}
%\pacs{05.40.Fb}{Random walks and Levy flights}
%\pacs{05.60.Cd}{Classical transport}
%

%\begin{document}

\maketitle 

\section{Introduction}

When two thermodynamic systems are placed in contact, the expected
result is that the systems equilibrate with one other.  In this
Letter, we show that the opposite may also occur: starting from
systems which individually equilibrate and placing them in contact
may instead drive the combined system \emph{out} of equilibrium,
leading to a \emph{nonequilibrium stationary state} (NESS) with a
nontrivial internal circulation flow. We show that this
counterintuitive behavior occurs already in one of the simplest
diffusive systems, biased random walkers, {where this effect is amenable to 
an exact analysis}.

It is  well known that a diffusive process subject to a constant drift, {or bias},
%directed 
towards a reflecting wall reaches an equilibrium state,
characterized by an exponential distribution decaying away from the
wall, known as the barometric formula \cite{perrin,chandra,wang, barrat}. 
%Precisely because this is an
{Being an}
equilibrium distribution, this can be interpreted as the configuration
factor of a Boltzmann distribution,
where the constant bias is due to
a force derived from a linear potential, {e.g., gravity, or an electric field}, 
and the diffusion constant is related to the temperature through the Einstein relation \cite{perrin,bringuier}. 
%Agregar Van Kampen???
%
% These identifications, in the context of colloidal
% particles, are at the root of Perrin's experimental results, by which
{In particular, these results form the basis of the experimental results of Perrin showing the 
``objective reality of molecules'' via Brownian motion under gravity \cite{perrin}.}
% 
% ``\ldots the molecular theory of the Brownian movement can be regarded as
% experimentally established, and, at the same time, it becomes very
% difficult to deny the objective reality of molecules''  \cite{perrin}.
% The characteristic of such equilibrium states, distinguishing them
% from other stationary states, is that the current throughout the
% system is zero.

%The subject of the equilibrium state {for} diffusive systems in
% {The equilibrium state attained by diffusive systems in}
% constant field has  re-appeared in diverse contexts, frequently
% under the name of {``}barometric formula{''} \cite{perrin, barrat}. The
% characteristic of such equilibrium states, what distinguishes them
% from other steady states, is, of course, that the current throughout the
% system is zero.

On the other hand, the nature of nonequilibrium steady states has
attracted much attention; see, e.g., Ref.~\cite{qian}. Although the
statistical properties of these systems are independent of time,
currents driven by external forces are sustained within the systems and
detailed balance is not satisfied.

In this Letter, we study the effect of coupling together two 
biased diffusive systems with reflecting boundaries. Each of these
systems individually equilibrates, and coupling them at a single site produces a global equilibrium state.
 However, when the
%on its own reaches an equilibrium state; however, when the
systems are coupled appropriately, we show that the combined system in
general reaches a nonequilibrium steady state (NESS) with a
non-vanishing current,
i.e., % In such a state, there is%, in effect, 
a constant particle circulation \cite{qian}, with particles traveling against the
force in one of the columns and returning through the other.
For definiteness, we present a {discrete-time model}, in which all walkers jump
at each time step; however,
our results can easily be transposed to a continuous-time version \cite{kac,weiss}.  
Results for periodic models have been obtained by Kolomeisky and co-workers \cite{kolomeisky_2001, kolomeisky_2006, kolomeisky_2013}.  %HOYLE?

\section{Model and equilibrium states}
\subsection{Model}
\begin{figure}
\includegraphics[scale=0.5]{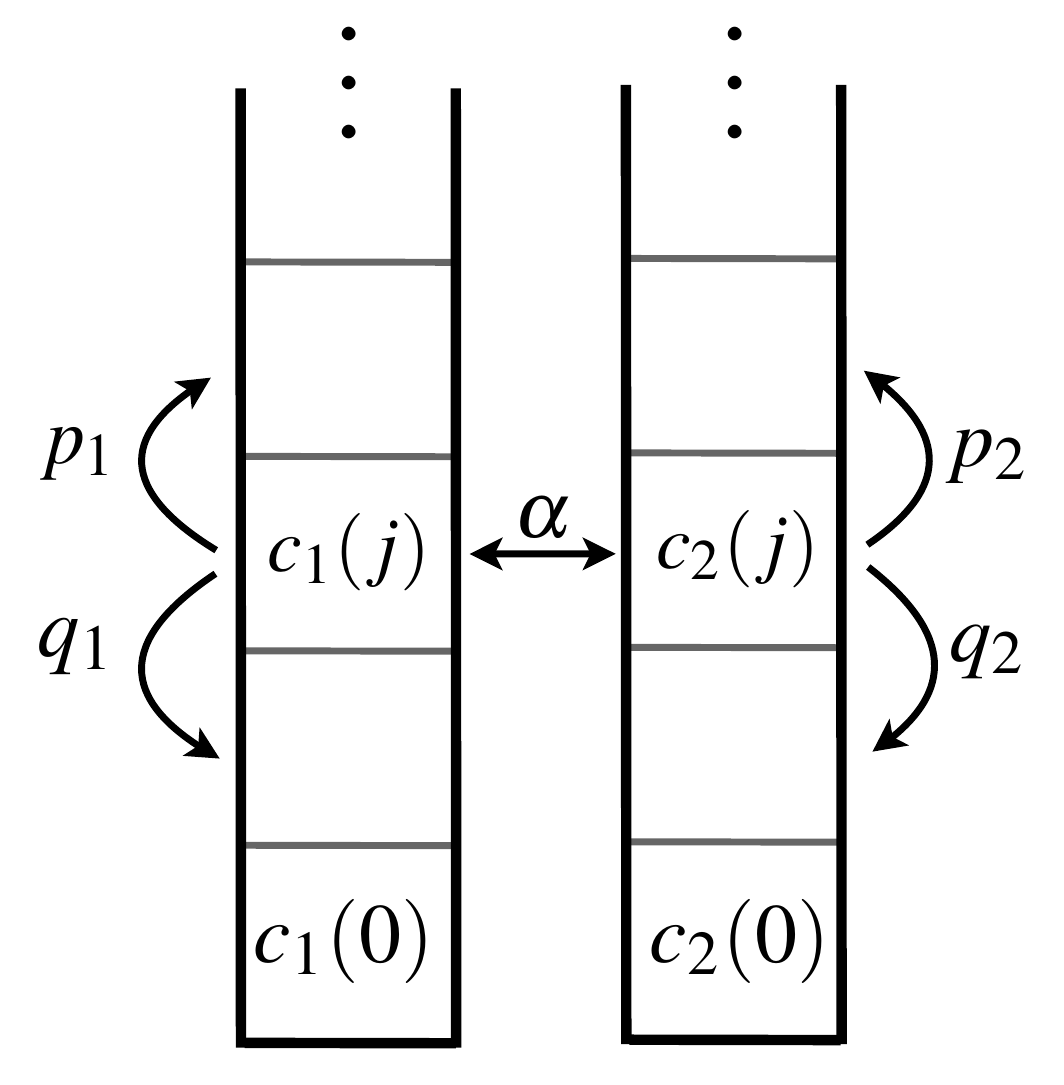}
\caption{Transition probabilities between
  sites and stationary occupation probabilities.
\label{fig:sketch}
}
\end{figure}

 Our model consists of two
vertical, semi-infinite discrete columns, with sites at height $i$
having concentrations $\cc{1}{i}{n}$ and $\cc{2}{i}{n}$, respectively,
at time step $n$; see Fig.~\ref{fig:sketch}.  Reflecting
boundaries {are placed} at the bottom of each column, which we take to be at $i=0$.
In each column $\nu=1,2$, we consider biased random walkers in
discrete time, with probabilities $\tp_\nu$ to jump upwards and
$\tq_\nu=1-\tp_\nu$ to jump downwards at each time step. To enforce
the reflecting boundary condition, walkers which try to jump
downwards from site $0$ remain at site $0$. 
We believe that this is the
simplest model in which the effect is found, allowing a complete analytical solution.
However, as we discuss below, similar phenomena can be expected to arise whenever two
or more systems with spatially varying equilibrium states are coupled
at several places.

When the columns are not coupled, the following master equation
describes the time evolution of the concentrations: %at each site:
\begin{equation}
 c_\nu(i ,n+1) = \tp_\nu c_\nu(i-1,n) + \tq_\nu c_\nu(i+1,n),  \quad j \neq 0
\end{equation}
with boundary conditions
\begin{equation}
c_\nu(0,n+1) = \tq_\nu c_\nu(0,n) + \tq_\nu c_\nu(1,n).
\end{equation}
If $\tq_\nu \geq \tp_\nu$, i.e., if the bias is directed towards the
reflecting boundary at the origin, then the physical (i.e.,
normalizable) stationary solutions of these equations are:
\begin{equation}
c_\nu(i) = {K}_\nu \left(\tp_\nu/\tq_\nu\right)^i,
\end{equation}
where the ${K}_\nu$ are normalization constants. The crucial feature of
these solutions is that the {vertical} current $J_\nu(i) \defeq \tp_\nu c_\nu(i-1)-\tq_\nu c_\nu(i)$ between sites within {either} column
vanishes
identically, so that the system is in \emph{equilibrium}.

\subsection{Coupling}
We now couple the columns together so that
%Now consider the case in which we 
%allow the columns to
they may exchange particles
 at a \emph{single} height %site 
$j$, % Specifically, assume that at site $j$,
where particles have the same probability $\alpha$ to jump ``horizontally'' from
one column to the other in either direction (see
Fig.~\ref{fig:sketch}). The equations describing
the system are then modified in the vicinity of site $j$ to take account of
the additional current due to the coupling:
\begin{eqnarray}
 c_\nu(j+1,n+1) = (1-\alpha)\tp_\nu c_\nu(j,n) + \tq_\nu c_\nu(j+2,n);
\nonumber
 \\
 c_\nu(j-1,n+1) = \tp_\nu c_\nu(j-2,n) + (1-\alpha)\tq_\nu c_\nu(j,n);   \nonumber \\
 c_1(j,n+1) = \tp_1 c_1(j-1,n) + \tq_1 c_1(j+1,n) +\alpha c_2(j,n);
\nonumber
 \\
 c_2(j,n+1) = \tp_2 c_2(j-1,n) + \tq_2 c_2(j+1,n) +\alpha c_1(j,n).
\nonumber
\end{eqnarray}
There is then a \emph{unique} stationary solution for the whole system, given by
% The stationary solution for the system in this case is given by
\begin{equation}
 c_\nu(i)=K \cases{ (1-\alpha)\left(\tp_\nu/\tq_\nu\right)^{i-j} & if  $i\neq j$; \\
  1 & if  $i=j$,}
\end{equation}
% where now there is a single normalization constant $N$ as a
for $\nu=1, 2$, with a single normalization constant $K$.
% consecuence of the coupling between the two columns. 
Note that the
concentrations in the two sites at height $j$, where interchange is
allowed, are equal, giving a zero total horizontal current there.  
Thus, this is again an equilibrium distribution
with vanishing current everywhere, showing that 
{such a coupling procedure is natural}, and 
%  This shows that there is nothing
% inherently problematic in coupling the two columns by allowing
% % particles to hop from one to the other with a fixed probability, and
% that doing so, 
allows the systems to equilibrate with each other.

\section{Non-equilibrium stationary state}

However, %we now show that 
coupling at \emph{more than one} site instead forces
the system to settle into a nonequilibrium stationary state, with non-vanishing currents.
%i.e., a non-equilibrium steady state.  
To show this, in what follows we consider the
case of horizontal coupling at every
height, with symmetrical probability $\alpha$ of jumping between columns.
%{PUT THE PHYSICAL INTERPRETATION HERE?}

In
% Again focusing only on 
the stationary state, the concentrations satisfy the following set of coupled linear equations at sites $i \neq 0$:
\begin{eqnarray} 
c_1(i) = \left(1-\alpha\right)\left[ \tp_1 c_1(i-1) + \tq_1
c_1(i+1) \right] + \alpha c_2(i), \label{eq:ci1} \label{eq:system-first}\\
c_2(i) = \left(1-\alpha\right) \left[ \tp_2 c_2(i-1) + \tq_2
c_2(i+1) \right] + \alpha c_1(i) \label{eq:ci2}.
 \end{eqnarray}
 At height $0$, the reflecting boundary conditions give the following:
\begin{eqnarray}
c_1(0) =  \left(1-\alpha\right)\left[\tq_1 c_1(0) + \tq_1 c_1(1)\right] + \alpha c_2(0) \label{eq:boundary_1},\\
c_2(0) =  \left(1-\alpha\right)\left[\tq_2 c_2(0) + \tq_2 c_2(1)\right] + \alpha c_1(0)\label{eq:boundary_2}.
\end{eqnarray}

To solve the system \eqref{eq:system-first}--\eqref{eq:boundary_2}, we introduce \emph{generating functions} \cite{weiss},
{a discrete version of the Laplace transform}:
\begin{equation}
 \Chat_\nu(z) \defeq \sum_{i=0}^{\infty} c_\nu(i)  z^i, \qquad \nu=1,2.
\label{eq:generating-fns}
\end{equation}
Multiplying \eqref{eq:ci1} and \eqref{eq:ci2} by $z^i$, summing
over $i$ from $1$ to $\infty$, and using the boundary conditions leads
to two simultaneous linear equations for the generating functions in
terms of the values of the concentrations $c_1(0)$ and
$c_2(0)$ at the boundary, which have yet to be determined:
\begin{equation}
\left(
\begin{array}{cc}
% \begin{pmatrix} 
f_1(z) & -\alpha\\
-\alpha & f_2(z)
% \end{pmatrix}
\end{array}
\right)
\left(
\begin{array}{c}
% \begin{pmatrix} 
\Chat_1(z) \\
\Chat_2(z)
% \end{pmatrix}
\end{array}
\right)
=
\left(
\begin{array}{c}
% \begin{pmatrix} 
g_1(z) \\
g_2(z)
% \end{pmatrix}
\end{array}
\right),
\end{equation}
where $f_\nu(z) \defeq 1 - (1-\alpha)[\tp_\nu z + \frac{\tq_\nu}{z}]$
and $g_\nu(z) \defeq (1-\alpha)\tq_\nu (1-\frac{1}{z}) c_\nu(0)$.

Solving yields
%\begin{widetext}
\begin{eqnarray}
\fl
\Chat_1(z)=\frac{ [z- (1-\alpha) (\tp_2z^2+\tq_2)] \tq_1 c_1(0) + \alpha z \tq_2
  c_2(0)} {\alpha z\left[ \tq_1+\tq_2 -z(\tp_1+\tp_2) \right]
  - (1-\alpha) (1-z) (\tq_1-z\tp_1) (\tq_2 -z\tp_2)},
\label{eq:chachan}
\end{eqnarray}
%\end{widetext}
and an analogous equation for $\Chat_2(z)$ with the subscripts $1$ and $2$ interchanged in the numerator (the denominator being unchanged).  
This gives (in principle) an exact solution for the stationary concentration distributions, obtained by inverting
Eq.~\eqref{eq:generating-fns}: $c_\nu(i) = \frac{1}{i!} \Chat^{(i)}_\nu(0)$, where $\Chat_\nu^{(i)}$ denotes the $i$th derivative; or, equivalently,
by expanding in formal power series in $z$ and equating coefficients.

The vertical current in column $\nu$ between site
$j$ and site $j+1$ is given by $J_\nu(j) \defeq (1-\alpha)[\tp_\nu
  c_\nu(j)-\tq_\nu c_\nu(j+1)]$.  The corresponding generating
functions follow from Eq.~\eqref{eq:chachan}: % and are given by
%\begin{widetext}
\begin{eqnarray}
\fl
\Jhat_1(z)=\alpha(1-\alpha) \frac{(\tq_2 -z\tp_2)\tq_1 c_1(0) -(\tq_1
    -z\tp_1) \tq_2 c_2(0)}{\alpha z\left[\tq_1+\tq_2 -z(\tp_1+\tp_2)\right]
  -(1-z)(1-\alpha)[\tq_1-z\tp_1][\tq_2 -z\tp_2]} 
\label{eq:chan}
\end{eqnarray}
%\end{widetext}
and $\Jhat_2(z) = - \Jhat_1(z)$, reflecting the fact that the
system is in a steady state. However, neither current (in either column)
can be identically zero unless the corresponding hopping probabilities
are equal, $p_1 = p_2$, in which case the steady state is an
equilibrium state. Rather, there is a circulation through the system,
moving upwards in the column with larger $p_{\nu}$ and downwards in
the other column.

\begin{figure}[t]
\includegraphics[scale=0.4]{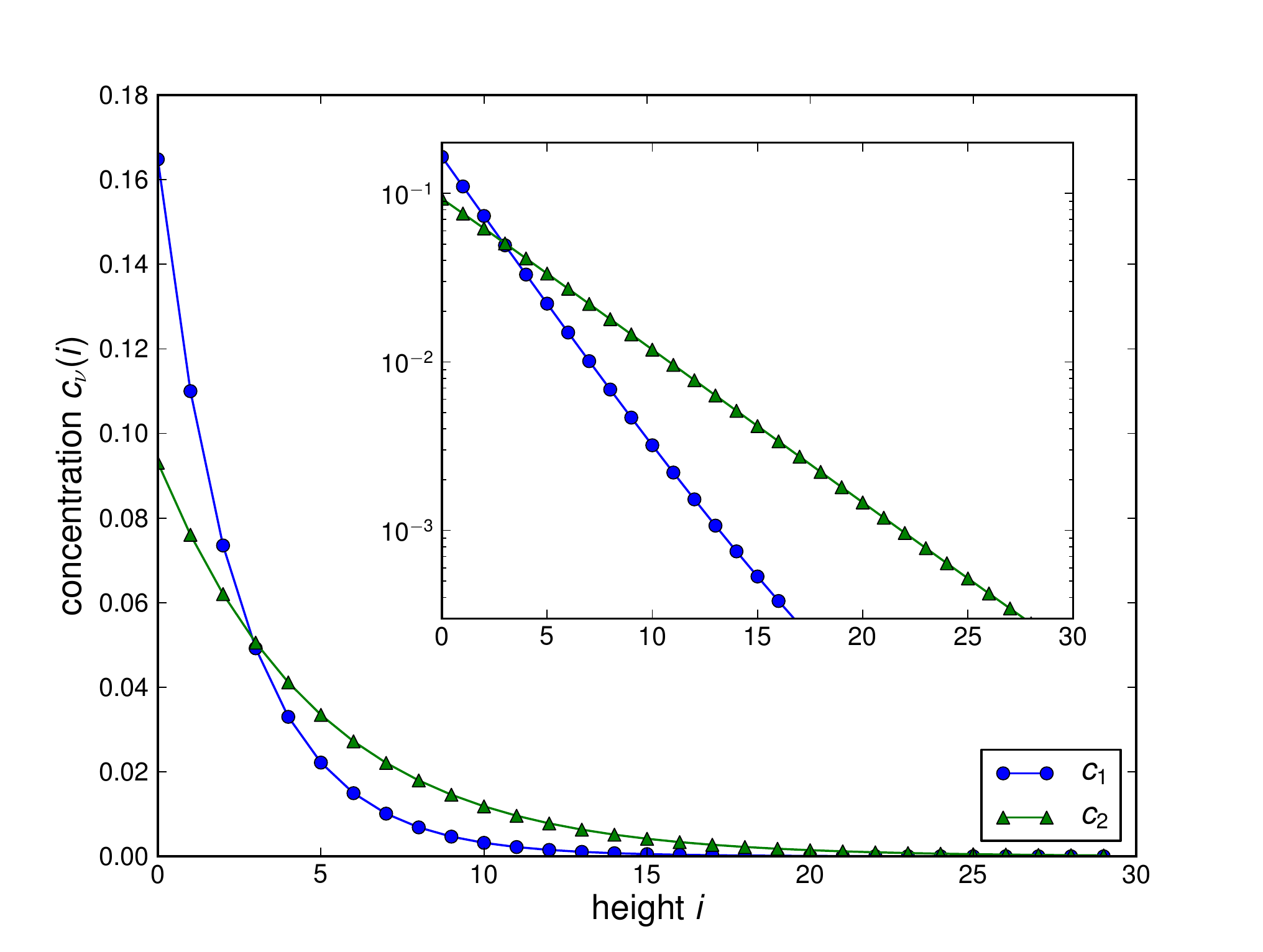}
\caption{Steady-state concentrations $c_\nu(i)$ as a function of
  height $i$ with coupling at all heights; the lines are a guide for the eye. Parameters: $p_1=0.4$; $p_2=0.45$; $\alpha=0.001$;
  system height $N=30$ with upper and lower reflecting boundary
  conditions. Inset: same data
  on a semi-logarithmic scale.
\label{fig:concs}
}
\end{figure}

\begin{figure}[t]
\includegraphics[scale=0.5]{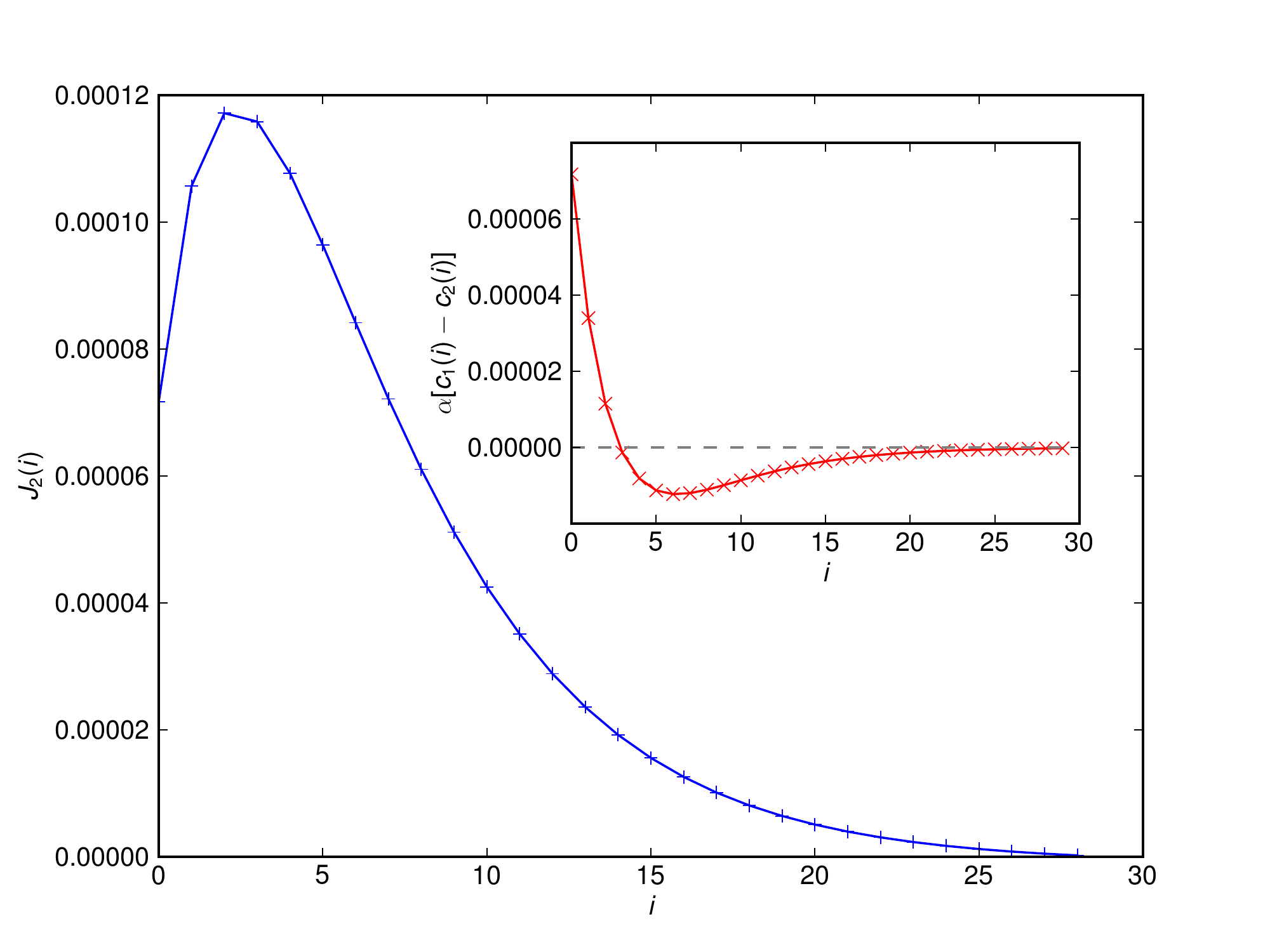}
\caption{Vertical current $J_{2}(i)$ as a function of height $i$; same
  parameters as in Fig.~\ref{fig:concs}. Inset: horizontal current as a function of height.
\label{fig:currents}
}
\end{figure}

Figures~\ref{fig:concs} and \ref{fig:currents} show this phenomenon for
a representative set of parameters, giving the concentration profiles,
and vertical and horizontal current, respectively. The stationary
solution was obtained by numerically iterating the master equation for
the time evolution of the $c_\nu(i)$, using a finite system with a reflecting boundary condition 
also at the top, confirming that the concentration 
profiles
% again 
have 
exponential tails. 
It may be checked that
% We can check that it agrees with the exact
this numerical solution agrees with the exact 
solution described below when the corresponding parameter values are 
substituted in the exact expressions.%, affected by finite-size effects
%for large $i$.  
% By modifying the parameters, the number of sites at
%  which the horizontal current is positive can be changed.

% \begin{figure}[tbp]
% \includegraphics[scale=0.4]{horiz_current}
% \caption{Horizontal current $\alpha [c_{1}(i) - c_{2}(i)]$ between sites
%   at height $i$, with the same parameters as in Fig.~\ref{fig:concs}.
% \label{fig:horiz_current}
% }
% \end{figure}

Other relevant quantities can also be easily calculated. For example,
since there is a sustained current in the system, it must be driven by
{thermodynamic} forces. %acting on the system. 
These forces can be related to the
respective hopping probabilities $\tp_\nu$ and $\tq_\nu$ in each
column via $f_\nu = \ln (\tp_\nu / \tq_\nu)$; this identification
arises from the fact that when $\alpha=0$, the concentrations reach
equilibrium distributions of the form $c_\nu(j) {\sim}
(\tp_\nu/\tq_\nu)^j=\exp [ - j \, \ln (\tq_\nu/\tp_\nu) ]$, 
%which
%justifies 
{so that} %identifying 
$j \, \ln(\tq_\nu/\tp_\nu)$ is a potential from
which the force $f_\nu$ is derived. Then, the amount of ``work''
dissipated can be calculated from the generating function
$\What(z)=\ln(\tp_1 /\tq_1) \Jhat_1(z)+\ln(\tp_2 /\tq_2) \Jhat_2(z)$,
giving that the total work dissipated within the system is
\begin{eqnarray}
\fl
\What(z=1) = (1-\alpha)\left[\ln(\tq_2 \tp_1/\tp_2 \tq_1)\right] 
  \frac{\left [(\tq_2 -\tp_2)\tq_1 c_1(0) -(\tq_1
    -\tp_1) \tq_2 c_2(0) \right]} {\left[ \tq_1+\tq_2 -(\tp_1+\tp_2) \right] }.
\end{eqnarray}

\section{Exact solution and asymptotics}
To evaluate and make use of {all of} the above {analytical} expressions, we must determine
the hitherto unknown constants $c_1(0)$ and $c_2(0)$. Taking the limit
$z \to1$ {in Eq.~\eqref{eq:chachan} gives} $\lim_{z \to 1} \Chat_1(z) = \lim_{z \to 1}
\Chat_2(z)$, i.e., the total probabilities in each column are equal in
the stationary state.  This is a result of the nature of the coupling {used},
and expresses the fact that the \emph{total} current from one column to
the other must be $0$ in the stationary state. (Note that the
(horizontal) current from one column to the other, at height $i$, is
$\alpha [c_1(i) - c_2(i)]$.)

In what follows, we assume the total normalization to be equal to one,
so that $\Chat_1(z=1) = \Chat_2(z=1) = \frac{1}{2}$.  Evaluating this equality gives
one relation between the unknown constants $c_1(0)$ and $c_2(0)$:
\begin{equation}
\tq_1 c_1(0)+ \tq_2 c_2(0)= 1 - (p_1 + p_2) = \tq_1 - \tp_2.
\end{equation}
To determine another relation between them, and hence find their precise values, we use the face that the generating
functions $\Chat_{\nu}$  be analytic in the region $|z|\leq 1$ \cite{satya}.
Indeed, $c_\nu(j)$ are probabilities, and thus take
values in the interval $[0,1]$, %in particular, that the
{so that} 
generating functions $\Chat_\nu(z)$ can have %singularities 
no {poles} in the
interval $z\in [0,1]$.

To impose this constraint, we must determine the zeros of the
denominator $\Delta(z)$ in the expression \eqref{eq:chachan} for the $\Chat_{\nu}(z)$, given by
\begin{eqnarray}
\fl
 \Delta(z) \defeq 
\alpha z\left[\tq_1+\tq_2 -z(\tp_1+\tp_2)\right] 
   -(1-z)(1-\alpha)(\tq_1-z\tp_1)(\tq_2 -z\tp_2).
\end{eqnarray}
 Since
$\Delta(0) = - \tq_1 \tq_2 (1-\alpha) < 0$ and $\Delta(1) = 2 \alpha
(\tq_1 + \tq_2 - 1)$, a sufficient condition for $\Delta(z)$ to have a
zero in the interval $(0,1)$ is that $\tq_1 + \tq_2 > 1$. We denote
the zeros of $\Delta$ by $z_0 \in (0,1)$ and $z_1, z_2 \ge 1$.  For
$\Chat_1(z)$ and $\Chat_2(z)$ to have no poles in $(0,1)$, the numerator in
Eq.~(\ref{eq:chachan}) must thus also vanish at $z_0$.

Since $\Delta(z)$ is a cubic polynomial, one can find exact explicit
expressions for $z_0$ in terms of the parameters of the
system. However, the general resulting expressions are cumbersome and
unenlightening, and we do not give them here.

 Nonetheless, simple perturbative expressions can be
found, for example, in the limit of small $\alpha$, for which we have, to first order in $\alpha$,
\begin{equation}
z_0 \simeq 1-\alpha\frac{(\tq_1+\tq_2)-(\tp_1+\tp_2)}{(\tq_1-\tp_1)(\tq_2-\tp_2)},
\end{equation}
provided $\tq_1 > \tp_1$ and $\tq_2 > \tp_2$. If, instead, $\tp_1 >
\tq_1$, say, and $\tq_2 $ is large enough to ensure that $\Delta(1) >
0$, then to first order %in $\alpha$ we have
\begin{equation}
z_0 \simeq \frac{\tq_1}{\tp_1}-\alpha\frac{\tq_1}{\tp_1-\tq_1},
\end{equation}
whereas if $\tq_1 = \tp_1$ then
% \begin{equation}
$z_0 \simeq 1-\sqrt{2\alpha}$.
% \end{equation}

Imposing the condition of analyticity on, say, $\Chat_1(z)$ then provides the
missing condition to determine $c_1(0)$ and $c_2(0)$:
\begin{equation}
[z_0-(1-\alpha)(\tp_2z_0^2+\tq_2)]\tq_1 c_1(0) + \alpha \tq_2  c_2(0)=0.
\end{equation}
Note that, given the form of the denominator, had we chosen instead to
impose the requirement of analyticity at $z_0$ on $\Chat_2(z)$, we would
have arrived at the same condition. Equivalently, we could impose that
the generating function for the current, Eq.~(\ref{eq:chan}), be analytic
at $z_0$. This leads to
\begin{equation}
c_1(0)=\frac{1}{2}\left[\frac{(\tq_1+\tq_2)-(\tp_1+\tp_2)}{(\tq_1+\tq_2)-z_0
    (\tp_1+\tp_2)}\right](\tq_1-z_0\tp_1) 
\end{equation}
and a corresponding equation for $c_2(0)$, which provide the complete
solution to the problem in terms of the root $z_0$. Then, for example,
substitution of these expressions in the equation for the total work
dissipated in the system gives
\begin{equation}
\What(z=1)=
\frac{\alpha z_0}{2}
\left[\frac{\tp_1\tq_2-\tp_2\tq_1}{(\tq_1-z_0\tp_1)(\tq_2
    -z_0\tp_2)}\right] 
\ln \left ( \frac{\tq_2\tp_1}{\tq_1\tp_2} \right).\nonumber
\end{equation}

The generating functions $\Chat_\nu(z)$ of the concentrations, as shown in
Eq.~\eqref{eq:chachan}, are rational functions of $z$, which
correspond to concentration profiles formed by a linear combination of
decaying exponentials $z_1^{-j}$ and $z_2^{-j}$. As is the case for
$z_0$, approximate expressions for these other roots at small values of
$\alpha$ can easily be calculated. For example, when $\tp_1 < \tq_1$
and $\tp_2 < \tq_2$,  to first order in $\alpha$ we have
\begin{equation}
z_\nu \simeq
\frac{\tq_\nu}{\tp_\nu}+\alpha\frac{\tq_\nu}{\tq_\nu-\tp_\nu},\qquad \nu=1,2.
\end{equation}

A special case occurs when one of the $\tp_\nu$ vanishes, say $p_{2} =
0$. In this case, the cubic denominator $\Delta(z)$ simplifies to a
quadratic, and the exact expressions for its zeros become
significantly simpler:
\begin{equation}
%z_{1,2} = \frac{1+\alpha - p_{1} \alpha -\sqrt{(-1-\alpha +p_{1} \alpha )^2 - 4 p_{1} (1-p_{1}-\alpha + p_{1} \alpha )} }{2 p_{1}}
z_{0,1} = \frac{1+\alpha q_{1}  \pm \sqrt{(1+\alpha q_{1} )^2 - 4 p_{1} q_{1} 
(1-\alpha) } }{2 p_{1}},
\end{equation}
where $z_{0} \in [0,1]$ is given by taking the minus sign for the
radical.  The $\Chat_{\nu}(z)$ must vanish at $z=z_{0}$, giving a single
factor $z-z_{1}$ in the denominator.  The two concentration profiles
are then exactly exponential for $i>0$, differing %from one another
only by a multiplicative factor and the concentration at
height $0$. 

\section{Conclusions}

\begin{figure}[t]
\includegraphics[scale=0.5]{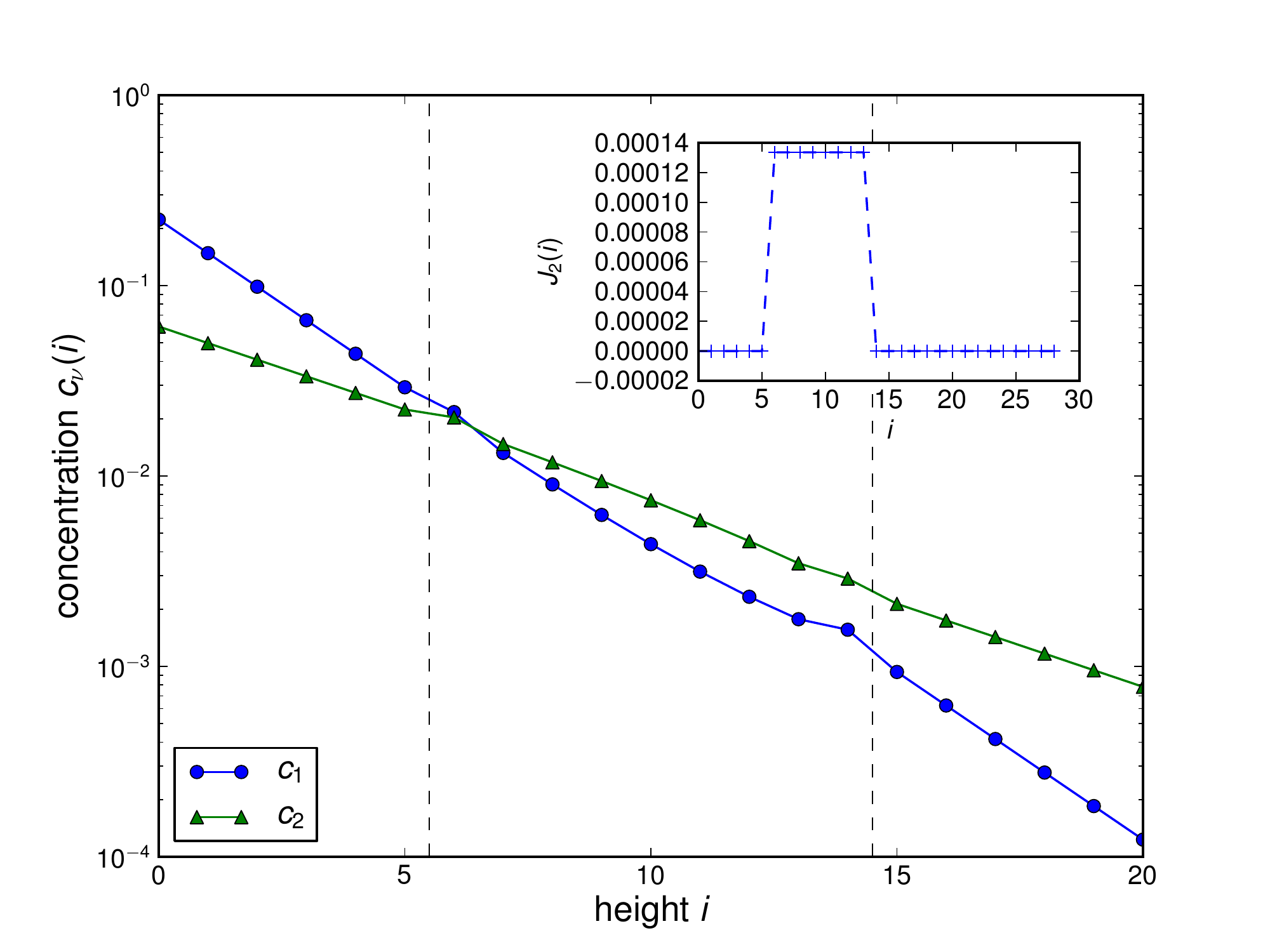}
\caption{Concentration profiles $c_\nu(i)$ (shown on a semi-logarithmic scale) as a function of height $i$ when the horizontal coupling is allowed only at two 
heights,
$j_1 = 6$ and $j_2=14$; vertical dashed lines indicate the different regimes. Parameters: $p_1=0.4$, $p_2=0.45$, $\alpha=0.1$, $N=30$.
Inset: Vertical current $J_2(i)$, which is constant between $j_1$ and $j_2$, as it must be.
\label{fig:two-site-coupling}
}
\end{figure}

In conclusion, by furnishing a complete analytical solution, we have shown that
coupling together two simple biased diffusive systems can give rise to
nonequilibrium steady state with a sustained current throughout the
combined system, even though each of them separately reaches
equilibrium. This effect arises due to the fact that the equilibrium
profiles can be ``matched'' when the systems are coupled at just a
single site, so that the current across that site vanishes, whereas if
the coupling occurs at various sites, then the equilibrium distributions
can no longer be made to match at all sites simultaneously.

The same phenomenon thus occurs even with coupling at only two
heights, $j_1$ and $j_2$. The density profiles and currents obtained numerically for a
specific realization of this minimal case are shown in Fig.~\ref{fig:two-site-coupling}.
This case can also be solved analytically: %turns out to be
% {is}
% rather messy. 
below $j_1$ and above $j_2$, the profiles are exactly exponential, due to the absence of current there;
these exponentials are   matched with the exact solution between $j_1$ and $j_2$.
% Another interesting variation, currently under study,
% consists in coupling three biased diffusive systems, say $c_1$, $c_2$
% and $c_3$, to see whether it is possible to sustain a ``winding''
% current which flows, say, $c_1 \to c_2 \to c_3 \to c_1 \to \cdots$.

Physical realizations of this phenomenon may be obtained, for example,
with particles suspended in a fluid under the combined effect of gravity and an applied
electric field, as in electrophoresis.  If the particles are capable
of capturing or losing %detaching 
a charge (or can be induced to do so), 
%that 
then they can %``jump'' 
transition 
from one charge state to the other. % then the
The drift induced by the field will then differ depending on the charge state of the
particle. Note that our model can also be interpreted in this way, where the subindex
describes the charge state of the particle, rather than its physical position.

These and other realizations may require generalizing the
analysis to the case in which there is asymmetric coupling, i.e., in which %That is,
the probability of jumping from state 1 to state 2 differs from the reverse transition probability.
%the probabilty of jumping from 2 to 1. 
The effect described in this work will nonetheless 
persist.
%  and the system will behave in a way that resembles the
% phenomenon we have described.

Financial support from SEP-CONACYT grant CB-101246 and PAPIIT 
grants IN109111, IN116212 and IN117214 from DGAPA-UNAM is acknowledged.

\section*{Bibliography}

\bibliographystyle{unsrt}
\bibliography{noneq-walkers}

\end{document}